\pdfoutput=1
\documentclass[aps,prd,amsmath,floats,floatfix, twocolumn,
superscriptaddress,nofootinbib,showpacs]{revtex4-1}

\usepackage[T1]{fontenc}
\usepackage[utf8]{inputenc}
\usepackage{lmodern}
\usepackage{url}

\usepackage{verbatim}
\usepackage[utf8]{inputenc}
\usepackage[dvipsnames, usenames]{xcolor}
\definecolor{linkcolor}{rgb}{0.0,0.3,0.5}
\usepackage[hypertexnames=false, unicode, colorlinks=true, linkcolor=linkcolor,
citecolor=linkcolor, filecolor=linkcolor,urlcolor=linkcolor,
pdfusetitle]{hyperref}

\usepackage[all]{hypcap}
\usepackage{graphicx}
\usepackage{xspace}
\usepackage{amssymb}
\usepackage[normalem]{ulem} 
\usepackage{bm} 
\usepackage{bbm}

\usepackage{microtype}

\graphicspath{%
  {figs/}%
}

\DeclareMathAlphabet{\mathpzc}{OT1}{pzc}{m}{it}

\newcommand{\etal}{\textit{et al.\ }}

\newcommand{\rma}{(r-\alpha M)}
\newcommand{\adx}{(\bm{a}\cdot\bm{x}_\mh)}

\newcommand{\ks}{\text{KS}}
\newcommand{\hk}{\text{H}}
\newcommand{\kshs}{\text{KSHS}}
\newcommand{\mh}{\text{MH}}

\newcommand{\Mark}[1]{{\textcolor{BurntOrange}{{#1}}}}

\begin{document}

\title{Extending superposed harmonic initial data to higher spin}

\newcommand\caltech{\affiliation{TAPIR 350-17, California Institute of
Technology, 1200 E California Boulevard, Pasadena, CA 91125, USA}}
\newcommand{\cornell}{\affiliation{Cornell Center for Astrophysics
    and Planetary Science, Cornell University, Ithaca, New York 14853, USA}} 
\newcommand\cornellPhys{\affiliation{Department of Physics, Cornell
    University, Ithaca, New York 14853, USA}}

\author{Sizheng Ma}
\email{sma@caltech.edu}
\caltech

\author{Matthew Giesler}
\cornell
\caltech

\author{Mark A. Scheel}
\caltech

\author{Vijay Varma}
\thanks{Klarman fellow}
\cornellPhys
\cornell
\caltech

\hypersetup{pdfauthor={Ma et al.}}

\date{\today}

\begin{abstract}

Numerical simulations of binary black holes are accompanied by an initial
spurious burst  of gravitational radiation (called `junk radiation') caused by
a failure of the initial data to describe a snapshot of an inspiral that
started at an infinite time in the past. A previous study showed that the
superposed harmonic (SH) initial data gives rise to significantly smaller junk
radiation. However, it is difficult to construct SH initial data for black
holes with dimensionless spin $\chi\gtrsim0.7$. We here provide a class of
spatial coordinate transformations that extend SH to higher spin. The new
spatial coordinate system, which we refer to as superposed modified harmonic (SMH), is characterized by a continuous parameter ---
Kerr-Schild and harmonic spatial coordinates are only two special cases of this
new gauge. We compare SMH with the superposed Kerr-Schild
(SKS) initial data by evolving several binary black hole systems with
$\chi=0.8$ and $0.9$. We find that the new initial data still leads to less
junk radiation and only
small changes of black hole parameters (e.g. mass and
spin). We also find that the volume-weighted constraint violations for the new
initial data converge with resolution during the junk stage $(t\lesssim700M)$,
which means there are fewer high-frequency components in waveforms at outer
regions. 
\end{abstract}

\maketitle

\section{Introduction} 
\label{sec:introduction}


The detection of GW150914 \cite{Abbott:2016blz} and other binary compact objects
\cite{LIGOScientific:2018mvr,Abbott:2020tfl,190412,190425,190814} has opened a new era in astrophysics. With the
improvement of detector sensitivity, more and more events are expected to be
detected in the near future \cite{Abadie:2010cf}.  Therefore, an accurate
modeling of coalescing binaries is crucial for data analysis. Numerical
relativity (NR) remains the only {\it ab initio} method to simulate the
coalescence of binary black hole (BBH) systems. With NR, one can obtain the
entire BBH waveform including inspiral, merger, and ringdown. Moreover, gravitational
wave models~\cite{Varma:2019csw, Varma:2018mmi, Ossokine:2020kjp,
Cotesta:2018fcv, Bohe:2016gbl, Pratten:2020ceb, Garcia-Quiros:2020qpx,
Pratten:2020fqn} used to analyze detector data are ultimately calibrated
against NR.

Numerical simulations of BBHs are
based on splitting the Einstein equation into constraint and evolution parts,
where the constraint equations provide the initial data to evolve.
However, the constructed initial data does not exactly
correspond to a quasi-equilibrium state of an inspiral that started at an infinite
time in the past.
For example, the tidal distortion
of a BH is not fully recovered, and the initial data do not usually include gravitational radiation already present. As a result,
once the evolution begins, the system relaxes
into a quasi-equilibrium state, and gives rise to a pulse of spurious
radiation, which is referred to as `junk radiation'. Several attempts have
been made to reduce junk radiation, by introducing PN
corrections \cite{Alvi:1999cw,Yunes:2006iw,JohnsonMcDaniel:2009dq,Kelly:2009js,Reifenberger:2012yg,Tichy:2016vmv},
or by using a curved conformal
metric \cite{Lovelace:2008hd,JohnsonMcDaniel:2009dq,Varma:2018sqd}.


Recently, Varma \etal \cite{Varma:2018sqd} carried out a systematic
study of initial data and its effects on junk radiation and computational
efficiency of the subsequent time evolution.
The simulations studied in Varma \etal were performed with an NR code:
the Spectral Einstein Code (SpEC) \cite{spec}, where the construction of
initial data is based on the Extended Conformal Thin Sandwich (XCTS)
formulation \cite{York:1998hy,Pfeiffer:2002iy}. Within this formalism,
several free fields,
including the conformal metric, must be provided. 
Different choices of the free fields
generate different physical initial data; the data still correspond to two black holes with the same desired mass ratio and spins, but the initial tidal distortions and strong-field dynamics differ. 
Varma \etal showed that the
junk radiation and efficiency of the subsequent evolution depend on the given
free fields.  In particular, choosing the initial data based on two superposed
black holes in time-independent harmonic coordinates \cite{PhysRevD.56.4775}
(heretofore called superposed harmonic (SH) data) leads to less junk radiation
than superposed Kerr-Schild (SKS) initial data~\cite{Lovelace:2008hd},
which is typically used in SpEC simulations~\cite{Boyle:2019kee}.  Varma \etal
also found that SH initial data has higher computational efficiency.  However,
SH initial data works well only for BHs with dimensionless spin
$\chi\lesssim0.7$. For high-spin BHs, the horizons become so
highly deformed that it
is difficult to construct initial data (cf. Fig. 10 in
Ref.~\cite{Varma:2018sqd}).

For both SH and SKS initial data, the conformal spatial metric
and the trace of the extrinsic curvature are determined by superposing
the analytic solutions for two single Kerr black holes.  The
difference is that SKS uses the Kerr metric in Kerr-Schild
coordinates, and SH uses the Kerr metric in time-independent harmonic
coordinates~\cite{PhysRevD.56.4775}.  It may be surprising that making
a different coordinate choice --- the choice of coordinates for the
single-BH analytic solution --- leads to a different physical BBH
solution.  The reason is that the superposition of two single-BH
solutions does not solve the Einstein equations for a BBH and is used
to compute only \emph{some} of the fields; the remaining fields are
computed by solving constraints and by quasi-equilibrium conditions.
For a single black hole, following the complete
initial data procedure (including
solving the constraints numerically) for both SKS and SH would
result in the same physical Kerr metric but in different
coordinates.


In this paper, we
extend SH to higher spins by using a spatial coordinate map
to transform the free data for the single-BH conformal metric,
while retaining
harmonic time slicing for this single-BH conformal metric.
The coordinate transformation defines a class of spatial
coordinate systems that are characterized by a continuous parameter $\alpha$.
We refer to these coordinates as the modified harmonic (MH) coordinate
system.  MH coordinates are purely harmonic with $\alpha=1$ and
correspond to spatial KS when $\alpha=0$. Similar to the cases of SKS and SH, an initial data for a BBH system can also be constructed by superposing two single Kerr black holes in MH coordinates. We refer to this initial data as superposed modified harmonic (SMH).
For the BBH systems with $\chi>0.7$,
a value of $\alpha < 1$ results in less distorted horizons. However, it is
desirable to keep $\alpha$ as close to 1 as possible so that
SMH data still shares
the desirable properties of SH initial data. 

This paper is organized as follows. In Sec.\ \ref{sec:BBH-ID}, we provide some basic information about how we compute initial data and evolve BBH systems.
In Sec.\ \ref{sec:metric},
we compare the behavior of different
single-BH coordinate systems.
In particular, in Sec.\ \ref{sec:H-fail-reason} we explicitly
point out the numerical reason that SH does
not work for high-spin BHs.
This immediately leads to a class of spatial coordinate transformations,
defined in Sec.\ \ref{sec:way-to-fix}, that
can cure the numerical issues.
We then use the MH coordinate system to construct initial data
for BBHs (i.e., SMH) with $\chi=0.8$ and $0.9$ and evolve these systems.
In Sec.\ \ref{sec:results}, we discuss the results of our simulations.
Finally in Sec.\ \ref{sec:conclusion}, we discuss our results and
highlight possible future
work.

Throughout the paper, we use Latin letters to stand for the spatial indices, and use Greek letters to represent spacetime indices.

\section{BBH initial data and evolution} 
\label{sec:BBH-ID}

Following the discussions in Ref.\ \cite{Varma:2018sqd}, we use the
XCTS formulation to construct initial data for a binary black
hole system.
Within this formalism,
one can freely specify the conformal metric $\bar{g}_{ij}$,
trace of extrinsic curvature $K$, and their time derivatives
$\partial_t\bar{g}_{ij}$ and $\partial_tK$.
To obtain quasi-equilibrium initial data, we choose
\begin{align}
\partial_t\bar{g}_{ij}=0, \quad \partial_tK=0 \, .
\end{align}
The construction of  the
other free fields, $\bar{g}_{ij}$ and $K$, is based on the 3-metric $g_{ij}^\beta$ and the trace of extrinsic curvature $K^\beta$ of
two single boosted Kerr BHs , where the superscript $\beta=1,2$ labels each of the two BHs in the binary system. The conformal metric and the trace of the
extrinsic curvature are then given by:
\begin{align}
&\bar{g}_{ij}=f_{ij}+\sum_{\beta=1}^{2}e^{-r_\beta^2/w_\beta^2}(g_{ij}^\beta-f_{ij}),\\
&K=\sum_{\beta=1}^{2}e^{-r_\beta^2/w_\beta^2}K^\beta \, ,
\end{align}
where  $f_{ij}$ is the flat 3-metric,
and $r_{\beta}$ is the Euclidean coordinate distance from
the center of each BH~\cite{Lovelace:2008tw}.
Note that each metric is weighted by a Gaussian with width
\begin{align}
w_\beta=0.6 \, d_\beta^{L_1} \, ,
\end{align}
where $d_\beta^{L_1}$ is the Euclidean
distance between the Newtonian
$L_1$ Lagrange point and the
center of the black hole labeled by $\beta$.
Here $g_{ij}^\beta$ and $K^\beta$ correspond to the Kerr solution
expressed either in the KS, harmonic, or MH coordinate systems. BBH initial data constructed from
the two Kerr solutions in the aforementioned coordinates are referred to as SKS, SH,
and superposed modified harmonic (SMH), respectively.

After specifying the free fields, the initial data are
completed by solving a set of coupled elliptic equations that ensure
satisfaction of the
constraints and an additional quasi-equilibrium condition.
Additionally, these elliptic equations require boundary conditions. At the outer boundary (typically chosen to be $10^9 \, M$ from the sources),
we impose asymptotic flatness  [cf. Eq.\ (11)---(13) in Ref.\
\cite{Varma:2018sqd}], and at each inner boundary we enforce an apparent horizon
condition [cf.~Eq.~(15)---(24) in Ref.\ \cite{Varma:2018sqd}]. 
After generating initial data in the XCTS formalism, we also need to specify the initial gauge for time evolution. Here we use the most common choice for SpEC simulations: $\partial_tN=\partial_tN^i=0$ in a corotating frame, where $N$ is the lapse function
and $N^i$ is the shift vector.
It was shown that the damped harmonic gauge \cite{Szilagyi:2009qz} is the
most suitable for mergers, so we do a smooth gauge transformation on a time scale of $\sim50 \,M$
during the early inspiral, to transform from the initial gauge to the better
suited damped harmonic gauge.

\section{Modified harmonic coordinate system} 
\label{sec:metric}
In this section, we aim to investigate the reason that makes the harmonic coordinates problematic for high-spin BHs. We begin with a brief review of KS coordinates in Sec.\ \ref{sec:KS-review}. Then in Sec.~\ref{sec:transformation-general}, we outline a method that can be used to study the numerical behavior of Kerr metric in different coordinate systems. It is then applied to KS spatial coordinates with harmonic slicing in Sec.~\ref{sec:kshs}, and to harmonic coordinates in Sec.~\ref{sec:hk}. Those analyses allow us to explicitly show the numerical
problem with using harmonic
coordinates
for high-spin BHs, as discussed in Sec.\ \ref{sec:H-fail-reason}. Finally in Sec.\ \ref{sec:way-to-fix}, we provide
a coordinate map to fix the problem.

\subsection{Kerr in Kerr-Schild coordinates}
\label{sec:KS-review}
For a stationary Kerr BH with mass $M$ and angular momentum $\chi M^2$ in the $z$ direction, the metric in KS coordinates $x^{\mu}_{\ks} = (t_\ks,x_\ks,y_\ks,z_\ks)$ is given by \cite{Kerr:1963ud}
\begin{align}
ds^2=g_{\mu\nu} dx^\mu_\ks dx^\nu_\ks=(\eta_{\mu\nu}+2H l_\mu l_\nu)dx^\mu_\ks dx^\nu_\ks \, , \label{metric-ks}
\end{align}
where $\eta_{\mu\nu}$ is the Minkowski metric, $H$ is a scalar function, and $l_\mu$ is a null covariant vector. The expressions
for $H$ and $l_\mu$ are not used here but can be found in Ref.~\cite{Kerr:1963ud}. 
With KS coordinates, the radial Boyer-Lindquist coordinate $r$ can be written as \cite{Kerr:1963ud}
\begin{align}
&r^2=\frac{1}{2}(x_\ks^2+y_\ks^2+z_\ks^2-a^2) \notag \\
&+\left[\frac{1}{4}(x_\ks^2+y_\ks^2+z_\ks^2-a^2)^2+a^2z_\ks^2\right]^{1/2} \, , \label{r-x-ks}
\end{align}
or equivalently
\begin{align}
\frac{x_\ks^2+y_\ks^2}{r^2+a^2}+\frac{z_\ks^2}{r^2}=1 \, . \label{xyz-ks}
\end{align}
Here we have used $a=\chi M$ for the sake of conciseness.
The outer and inner horizons of the BH are located at
\begin{align}
r_\pm=M\pm\sqrt{M^2-a^2} \, . \label{hori-loc}
\end{align}
\subsection{Transforming from KS to a different coordinate system}
\label{sec:transformation-general}
Now we introduce a new coordinate system $x^\mu=(t,x,y,z)$, which are related to the KS coordinates $x^{\mu}_{\ks}$ through
\begin{align}
\left(\begin{matrix}
dt_\ks \\
dx_\ks \\
dy_\ks \\
dz_\ks
\end{matrix}\right)
=\left(\begin{matrix}
1 & \bm{b}\\
\bm{0} & \bm{C}
\end{matrix}\right)
\left(\begin{matrix}
dt \\
dx \\
dy \\
dz
\end{matrix}\right) \, , \label{coordinate-transform-general}
\end{align}
where $\bm{b}$ is a 3D vector, and $\bm{C}$ is a $3\times3$ matrix. In Eq.~(\ref{coordinate-transform-general}), we have assumed that the new spatial coordinates are independent of $t_\ks$\footnote{Equivalently, $(x_\ks,y_\ks,z_\ks)$ are independent of $t$.}. Note that we here keep the forms of $\bm{b}$ and $\bm{C}$ generic, so that our present discussion can be applied to different coordinate systems.

With the Jacobian at hand, we could transform the Kerr metric into the new coordinates, and study the numerical features of each metric component, such as the problematic behavior of harmonic
coordinates for high-spin black holes, but this usually involves very complicated calculations. However, since $g_{\mu\nu}^\ks$ can be decomposed into two pieces
[Eq.\ (\ref{metric-ks})], it is simpler to study the transformations of
$\eta_{\mu\nu}$\footnote{We have checked that the same problematic terms also occur
        in the $Hl_\mu l_\nu$ piece of
    Eq.~(\ref{metric-ks}). }.  In the new coordinates, we have 
\begin{align}
\eta_{\mu\nu}=&\left(\begin{matrix}
-1 & \bm{0} \\
\bm{0} & \mathbb{I}^3
\end{matrix}\right) \rightarrow  
\left(\begin{matrix}
1 & \bm{b} \\
\bm{0} & \bm{C}
\end{matrix}\right)^T
\left(\begin{matrix}
-1 & \bm{0} \\
\bm{0} & \mathbb{I}^3
\end{matrix}\right)\left(\begin{matrix}
1 & \bm{b}  \\
\bm{0} & \bm{C}
\end{matrix}\right) \notag \\
&=\left(\begin{matrix}
-1 & -\bm{b} \\
-\bm{b}^T & \bm{C}^T\bm{C}-\bm{b}^T\bm{b}
\end{matrix}\right) \, ,\label{eq:FlatMetricTransformation}
\end{align}
where $\mathbb{I}^3$ is the three-dimensional identity matrix. Both the 3-metric $(\bm{C}^T\bm{C}-\bm{b}^T\bm{b})$ and the shift vector $-\bm{b}$ above are modified by the vector $\bm{b}$.  
Any numerically problematic term in $\bm{b}$ might cause difficulty to resolve the metric in the new coordinates. 
Below, we focus on the $z$ component of $\bm{b}$, $b^z$, at the inner boundary $r=r_+$,
and study its numerical behavior for high-spin black holes (especially when $a\to M$) with several coordinates.

\subsection{Kerr-Schild spatial coordinates with harmonic slicing}
\label{sec:kshs}
We first apply our discussion in Sec.~\ref{sec:transformation-general} to a mixed coordinate system: KS {\it{spatial}} coordinates together with harmonic {\it{temporal}} slicing, then we have 
\begin{subequations}
\begin{align}
&\bm{C}_{\kshs}=\mathbb{I}^3, \\
&\bm{b}_{\kshs}=\frac{2M}{r-r_-}\bm{\nabla} r, \label{b-kshs}
\end{align}
\end{subequations}
where the subscript `KSHS' stands for Kerr-Schild spatial coordinates with Harmonic Slicing; and $r$ is the radial Boyer-Lindquist coordinate. Note that Eq.~(\ref{b-kshs}) is the result of \cite{PhysRevD.56.4775}
\begin{align}
t_\kshs=t_\ks-\int \frac{2M}{r-r_-} \, dr \, . \label{time-hk-ks}
\end{align}
We refer the reader to Appendix \ref{app:metric} for the detailed expression of $\bm{\nabla} r$. The $z$ component of $\bm{b}_{\kshs}$ at the inner boundary $r=r_+$ is given by (as $a\to M$)
\begin{align}
b^z_{\kshs}=\frac{M^2z_\ks}{r_+^4+(az_\ks)^2} \, . \label{U-ks}
\end{align}
\subsection{Harmonic coordinates}
\label{sec:hk}
Let us turn our attention to harmonic coordinates $x^{\mu}_{\hk} = (t_\hk,x_\hk,y_\hk,z_\hk)$, where the spatial coordinates also become harmonic. For such a coordinate system, we have \cite{PhysRevD.56.4775}
\begin{align}
&(r-M)^2=\frac{1}{2}(x_\hk^2+y_\hk^2+z_\hk^2-a^2) \notag \\
&+\left[\frac{1}{4}(x_\hk^2+y_\hk^2+z_\hk^2-a^2)^2+a^2z_\hk^2\right]^{1/2} \, , \label{r-x-hk}
\end{align}
and
\begin{align}
\frac{x_\hk^2+y_\hk^2}{(r-M)^2+a^2}+\frac{z_\hk^2}{(r-M)^2}=1 \, , \label{xyz-hk}
\end{align}
where the subscript `H' stands for harmonic coordinates.
The harmonic slicing implies 
\begin{align}
\bm{b}_{\hk}=\frac{2M}{r-r_-}\bm{\nabla} r,
\end{align}
with $z$ component of $\bm{b}_\hk$ at $r=r_+$ given by (as $a\to M$)
\begin{align}
b_\hk^z=\frac{M^2z_\hk}{(r_+-M)^4+(az_\hk)^2} \, . \label{U-hk}
\end{align}
Expressions for the
$3\times3$ block matrix, $(C_\hk)^i_{j}=\partial x_\ks^i/\partial x_\hk^j$
, along with additional details, can be found in Appendix \ref{app:metric}.

\newsavebox{\StrikeoutInsideSectionHeading}
\savebox{\StrikeoutInsideSectionHeading}{\Mark{s \sout{system}}}

\subsection{Problematic behavior of harmonic coordinates}
\label{sec:H-fail-reason}
\begin{figure}[htb]
        \includegraphics[width=\columnwidth,height=6.9cm,clip=true]{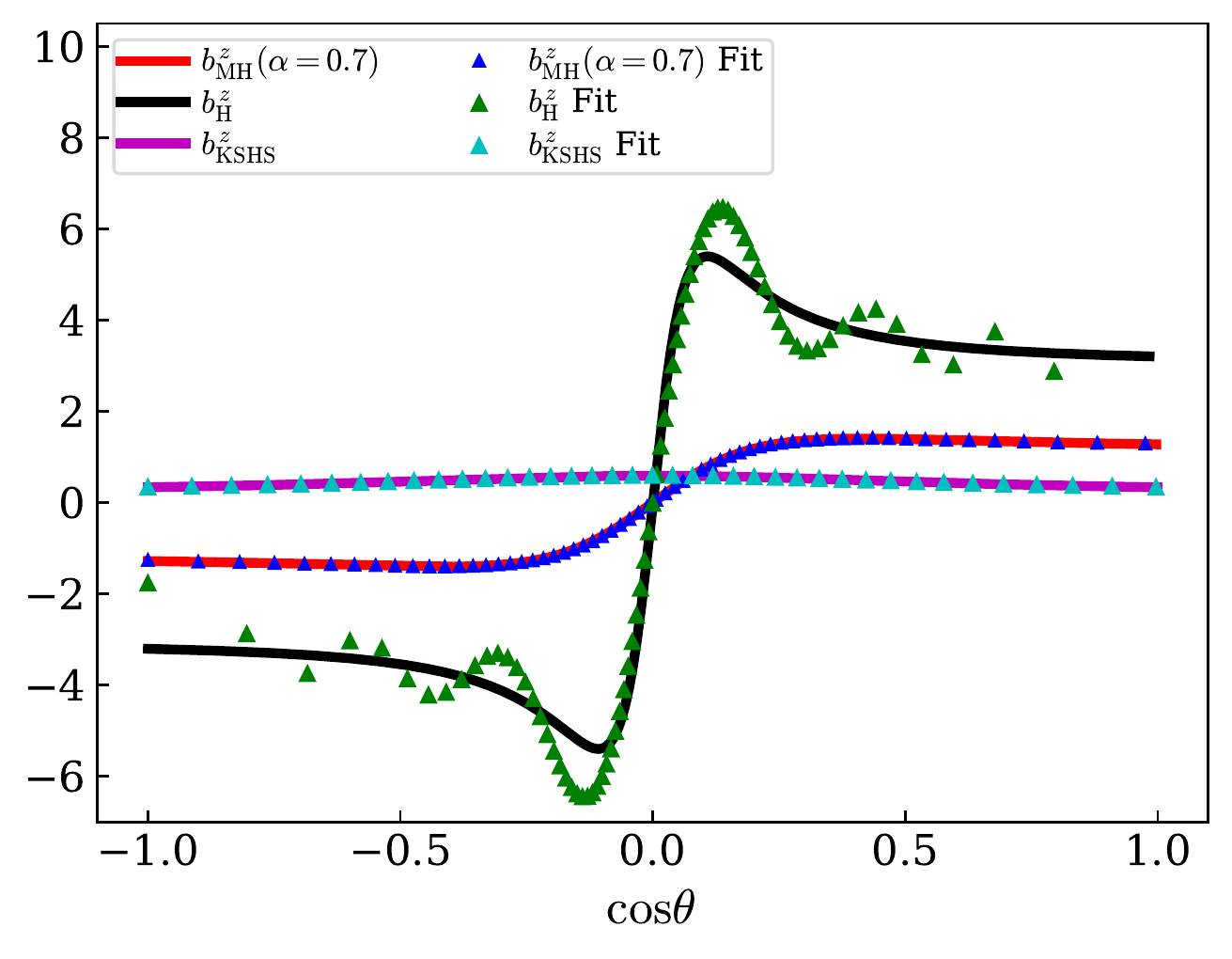} 
        \caption{The
        function
        $b^z$ in KSHS [Eq.\
            (\ref{U-ks})], harmonic [Eq.\ (\ref{U-hk})] and MH [Eq.\ (\ref{U-mh})] coordinates
            with $\alpha=0.7$.
            Solid lines represent
            $b^z$, whereas
            triangles represent
             the Legendre-Gauss collocation
            approximation to each function $b^z$ using $20$ Legendre polynomials.
            The spin
            of the BH is
            $a=0.95\,M$. $b^z$ is better approximated by a fixed
    number ($l=20$) of Legendre polynomials for MH than for harmonic coordinates.}
 \label{fig:dem-angular}
\end{figure}

In SpEC, the Legendre polynomials are used to numerically expand $b_\hk^z$ and $b_\kshs^z$ as functions of $\cos\theta$, defined by
\begin{align}
\cos\theta=\frac{z_\hk}{\sqrt{x^2_\hk+y^2_\hk+z^2_\hk}} \, . \notag 
\end{align}
 Here $\theta$ is the polar angle in harmonic coordinates and is not to be confused with the angular Boyer-Lindquist coordinate. As a test, we first represent $b_\hk^z$ [Eq.~(\ref{U-hk})] with twenty Legendre-Gauss collocation points and a BH spin of $a=0.95 \, M$. The results of this test are
shown in Fig.~\ref{fig:dem-angular}. From Fig.~\ref{fig:dem-angular}, we see that
the function $b_\hk^z$  is  difficult to
resolve using Legendre polynomials. This is
the primary reason that harmonic coordinates fail to accurately represent high-spin BH
initial data. Note that increasing the resolution to $l\gtrsim60$ (for a single BH)
eventually allows us to resolve $b_\hk^z$, but in practice requiring such
high resolution is computationally prohibitive; furthermore, the required
resolution increases rapidly as the spin increases.

Previous studies have shown success in high-spin BBH simulations with SKS initial data
up to spins of $\chi=0.998$ \cite{Scheel:2014ina}. 
A natural question to ask is whether the spatial or the time coordinates are more important in allowing KS coordinates to better resolve highly-spinning black holes.  Therefore, we also investigate the behavior of $b^z_\kshs$ (see Sec.~\ref{sec:kshs}) in which the time coordinate is harmonic but the spatial coordinates are Kerr-Schild.
Again, we represent $b^z_\kshs$ with twenty Legendre-Gauss collocation points and a BH spin of $a = 0.95M$, as shown in Fig.\ \ref{fig:dem-angular}.
The representation is much better than the case of harmonic coordinates. And we also confirm that with such mixed coordinates, BBH initial data can be indeed extended to higher spins. However, as we show later, they do not lead to
a smaller amount of junk radiation than SKS initial data.

Looking more closely at Fig.\ \ref{fig:dem-angular}, $b^z_\kshs$ has fewer
structures than $b_\hk^z$, which makes $b^z_\kshs$ easier
to represent by Legendre polynomials. More quantitatively, we write
\begin{align}
    b_\hk^z\sim\frac{1}{u^2+\frac{\epsilon}{1+\epsilon}}, \label{expand_bh}
\end{align}
with
\begin{align}
    u=\frac{a}{r_+-M}\cos\theta,\quad 
    \epsilon=\frac{(r_+-M)^2}{(r_+-M)^2+a^2}.
\end{align}
In Eq.~(\ref{expand_bh}), we have omitted unimportant functions of
$\cos\theta$ since they are well represented by Legendre
polynomials. We see $b_\hk^z$ has two poles $u=\pm
i\sqrt{\epsilon/(1+\epsilon)}$. The domain of convergence for Legendre
series $P_n(\cos\theta)$ is an elliptic region on the complex
plane \cite{boyd2001chebyshev}. If we restrict ourselves to the real
axis, we can obtain the radius of convergence as
\begin{align}
    |\cos\theta|\lesssim \sqrt{\frac{1+2\epsilon}{1+\epsilon}}\frac{r_+-M}{a}.
\end{align}
The radius becomes less than 1 if $a\gtrsim0.75M$, thus in that case
the Legendre
polynomials fail to provide a good representation for the metric. This
is the main reason that BBH simulations using SH become
difficult when spins are larger than about $a=0.7M$
\cite{Varma:2018sqd}. We remark that Chebyshev series
have the same domain of convergence as Legendre series; hence we do
not expect the situation can be improved by changing basis.

\subsection{Modified Harmonic coordinates}
\label{sec:way-to-fix}
We have seen that $b^z_\hk$ is sensitive to $\cos\theta$ for high-spin BHs. To reduce such dependence, we define a more general coordinate system
\begin{align}
\frac{x_\mh^2+y_\mh^2}{(r-\alpha M)^2+a^2}+\frac{z_\mh^2}{(r-\alpha M)^2}=1, \quad t_\mh=t_\hk \, ,
\end{align}
which leads to
\begin{align}
&(r-\alpha M)^2=\frac{1}{2}(x_\mh^2+y_\mh^2+z_\mh^2-a^2) \notag \\
&+\left[\frac{1}{4}(x_\mh^2+y_\mh^2+z_\mh^2-a^2)^2+a^2z_\mh^2\right]^{1/2} \, . \label{r-x-mh}
\end{align}
Here we introduce a new constant parameter $\alpha$.
As mentioned earlier, we refer to this new 
choice of spatial coordinates as the modified harmonic (MH) coordinate system.
MH coordinates
become harmonic ({\it{spatial}}) coordinates
when $\alpha=1$ [Eq.\ (\ref{xyz-hk})] and become KS ({\it{spatial}}) coordinates
when $\alpha=0$ [Eq.\ (\ref{xyz-ks})]. Meanwhile, the time slicing of MH coordinates is the same as in harmonic, regardless of the value of $\alpha$.
With this new coordinate system, the radius
of the outer horizon
along the spin direction is $(1-\alpha)M+\sqrt{M^2-a^2}$.
For $a\to M$, this radius goes to $M$ for KS coordinates
($\alpha=0$) and it goes to zero for harmonic coordinates ($\alpha=1$).
Therefore, the horizon with harmonic coordinates is highly compressed
in the spin direction. However, if we let $\alpha$ be a number smaller than,
but still close to 1, the horizon will be less distorted.
On the other hand, since $\alpha$ is close to 1, we can expect that
it still shares some similar properties (e.g. less junk radiation) with harmonic coordinates.  

\begin{figure}[htb]
        \includegraphics[width=\columnwidth,height=6.3cm,clip=true]{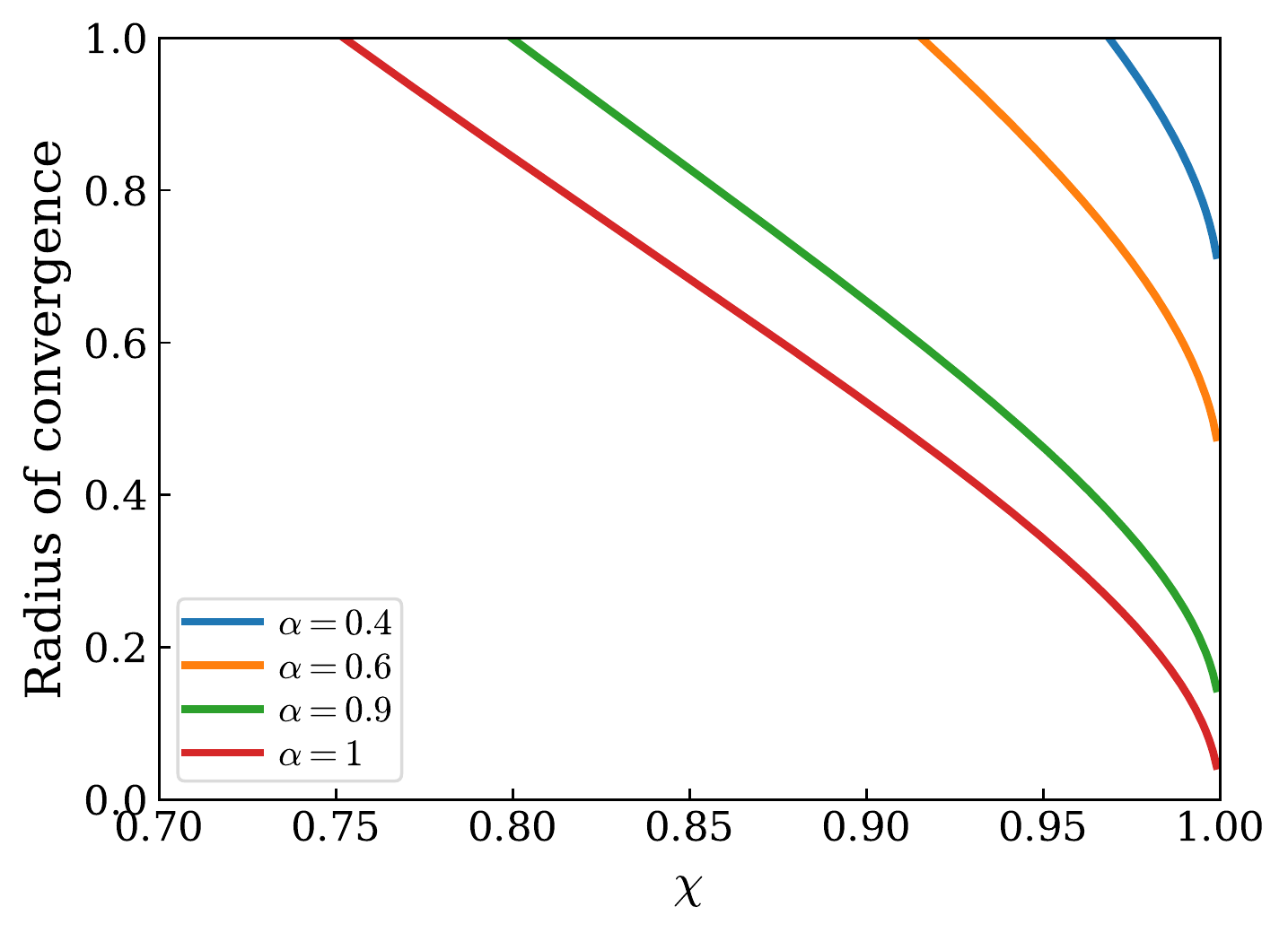} 
        \caption{The radius of convergence given in Eq.~(\ref{convergence_radius}).}
 \label{fig:convergence_radius}
\end{figure}

As in Sec.\ \ref{sec:H-fail-reason}, we use the function $b^z$ as an example
to see the improvement offered by MH coordinates.
In the MH coordinate system, we have
\begin{align}
b^z_\mh=\frac{M^2z_\mh}{(r_+-\alpha M)^4+(az_\mh)^2} \, . \label{U-mh}
\end{align}
Now $(r_+-M)^2$ is replaced by $(r_+-\alpha M)^2$.
The problematic part of $b^z_\mh$ takes the same form as
Eq.~(\ref{expand_bh}), except that 
\begin{align}
    u=\frac{a}{r_+-\alpha M}\cos\theta,\quad 
    \epsilon=\frac{(r_+-\alpha M)^2}{(r_+-\alpha M)^2+a^2}.
\end{align}
And the radius of convergence is given by 
\begin{align}
    |\cos\theta|\lesssim \sqrt{\frac{1+2\epsilon}{1+\epsilon}}\frac{r_+-\alpha M}{a}. \label{convergence_radius}
\end{align}
In Fig.~\ref{fig:convergence_radius}, we plot the radius of convergence as a function of $\chi=a/M$ for several values of $\alpha$. We see the convergent region for a fixed $\chi$ is enlarged if $\alpha$ becomes smaller.
As a consequence, it
should be easier for Legendre polynomials to represent $b^z_\mh$.
To see that this is the case, in Fig.~\ref{fig:convergence_radius} we
plot $b^z_\mh$ with $\alpha=0.7$ and $a=0.95 \,M$, using the same set
of angular Legendre-Gauss collocation points as for the other curves
in the figure.
As expected, the representation in Legendre polynomials of $b^z_\mh$ shows
an enormous improvement over the same representation of $b^z_\hk$.

\begin{figure}[htb]
        \includegraphics[width=\columnwidth,clip=true]{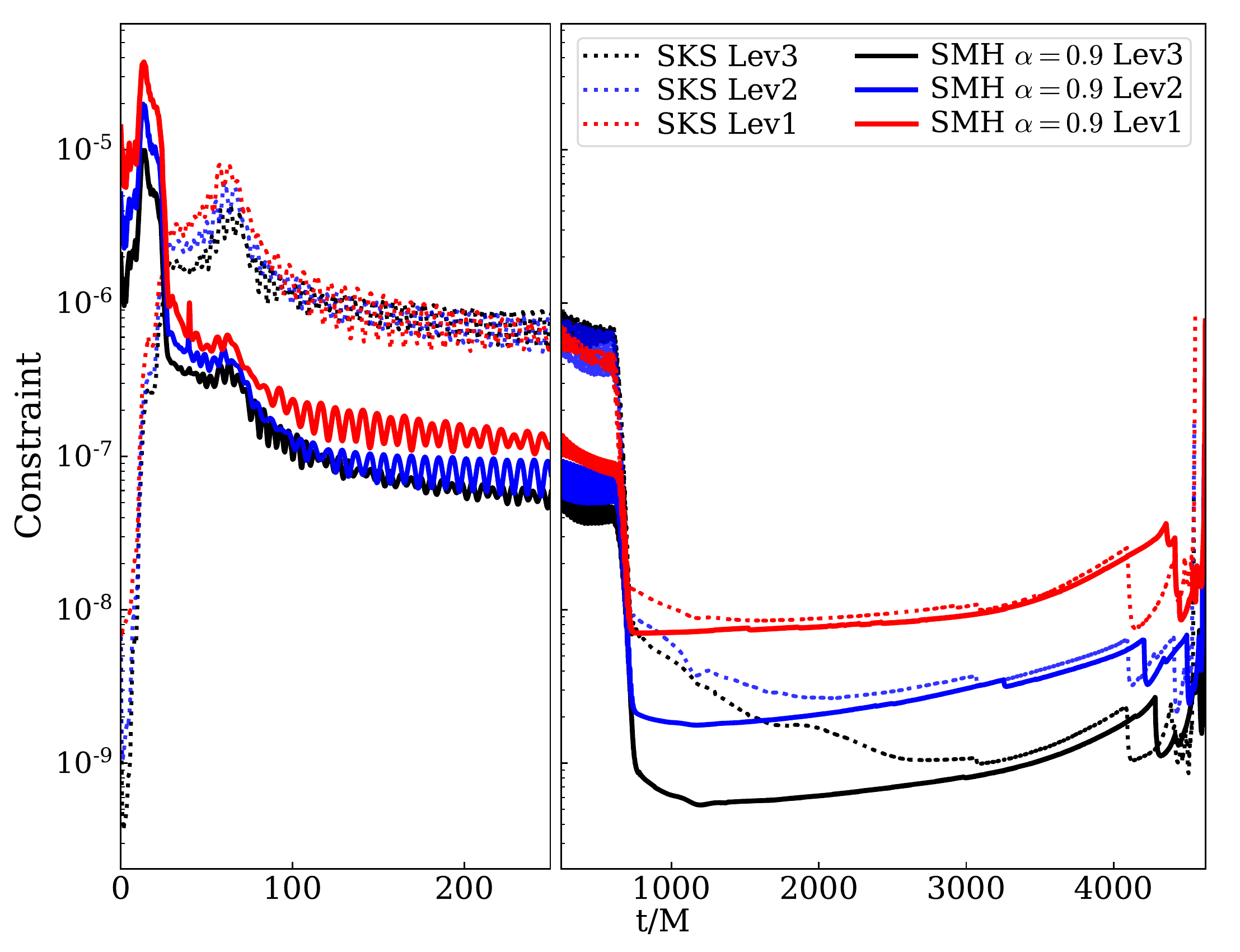} 
        \caption{ The volume-weighted generalized harmonic constraint energy
  for evolutions of Case I, with both SKS (dotted lines) and SMH (solid lines) initial data.
  Three resolutions are shown, labeled 'Lev1' (red), 'Lev2' (blue),
  and 'Lev3' (black) in order of decreasing AMR tolerance
  (i.e. in order of increasing numerical resolution).
  At the beginning,
  BHs of SMH initial data are more distorted on the grid
  so the constraints are worse.
  However, as the gauge transition proceeds, the constraints decay
  quickly.
  During most of the junk stage $(25M\lesssim t \lesssim700M)$, the constraints
  of SMH initial data are smaller than SKS by an order of magnitude.
  They also converge with resolution. After the junk stage,
  SKS and SMH finally become comparable.}
 \label{fig:constraint}
\end{figure}

\begin{figure}[htb]
        \includegraphics[width=\columnwidth,clip=true]{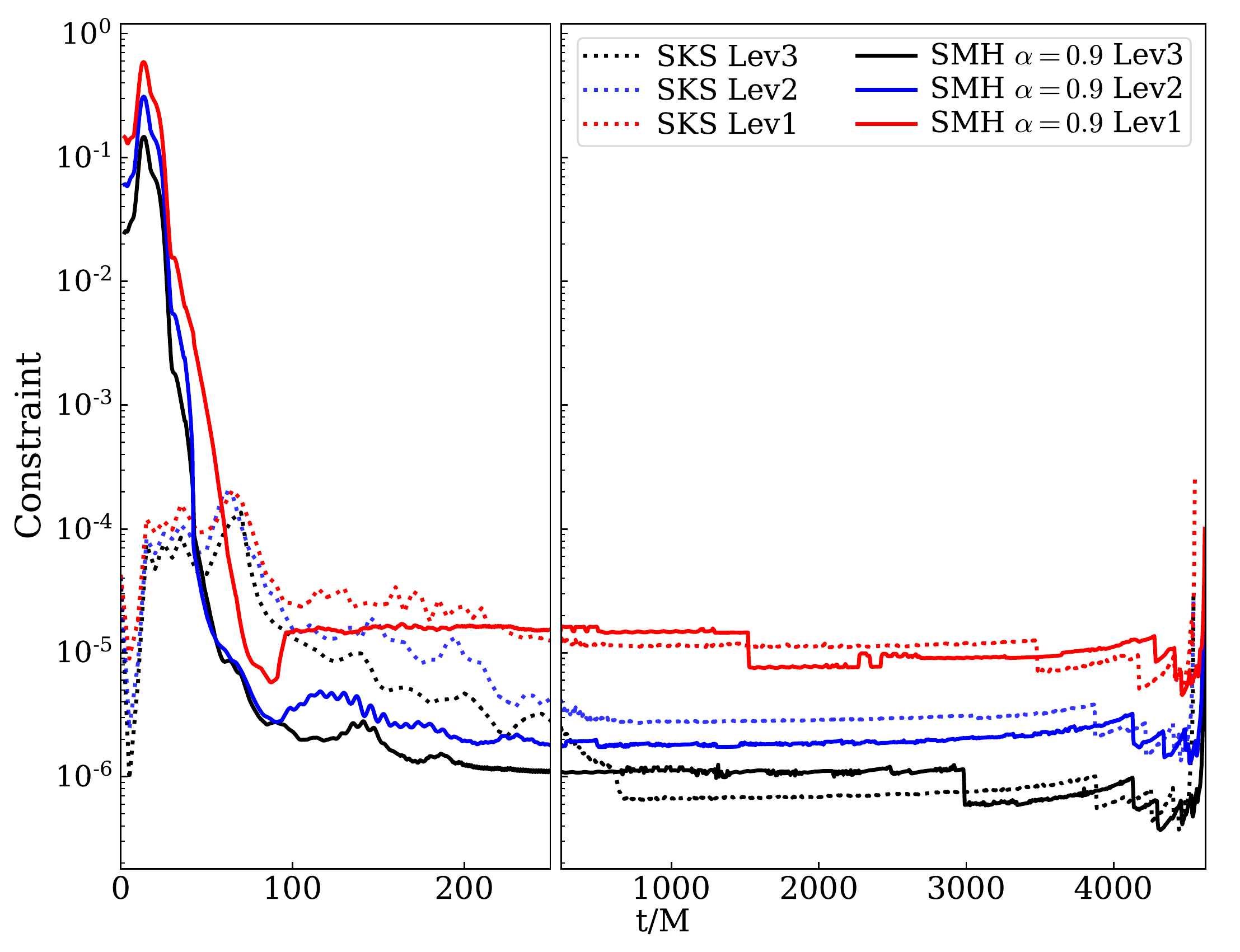} 
  \caption{Same as Fig.\ \ref{fig:constraint}, except that $L^2$ norm is used.}
 \label{fig:constraint_L2}
\end{figure}

\section{Results} 
\label{sec:results}
In this section we investigate the numerical behavior of BBHs evolved
starting with SMH initial data, compared to evolution of SKS data. We
pick four cases, as summarized in Table \ref{table:sim-params}. To
make comparisons, we consider constraint violations, computational
efficiency, changes of BH parameters (mass and spin), and junk
radiation. For the first three factors, we show the general features
of SMH by focusing on Case I. For junk radiation, we study all
cases. For each simulation, we evolve with three resolutions
(labeled Lev 1,2,3 in order of increasing resolution).
The resolution is chosen
by specifying different numerical error tolerances to the
adaptive mesh
refinement (AMR) algorithm \cite{Szilagyi:2014fna}.  The orbital
eccentricity is iteratively reduced to below
$\sim10^{-3}$ \cite{Buonanno:2010yk}.
The coordinate sizes of the black holes are different for SMH and SKS, so the excision boundaries (which are placed just inside each apparent horizon) are also different for SMH and SKS; this means that the grids are not exactly the same beween the two cases, but the grid points are chosen by AMR so that the two cases have the same approximate numerical error.

\begin{table}
    \centering
    \caption{A summary of parameters (mass ratio $q$ and dimensionless spins $\bm{\chi}$) for four simulations, where the spins of Case II are chosen randomly.
    The orbital
angular momentum is pointing along $(0,0,1)$. In the final column, we show the value of $\alpha$ for MH coordinates.}
    \begin{tabular}{c c c c c} \hline\hline
 Simulation label   & $q$ & $\bm{\chi}_{1}$ & $\bm{\chi}_{2}$ & $\alpha$ \\ \hline
Case I & 1 & $(0,0,0.8)$ & $(0,0,0.8)$& 0.9 \\ \hline
Case II & 1 & $(0.44,0.44,0.50)$ & $(0.13,0.64,0.46)$ & 0.9 \\ \hline
Case III & 2 & $(0,0,0.7)$ & $(0,0,0.8)$ & 0.8 \\ \hline
Case IV & 1 & $(0,0,0.9)$ & $(0,0,0.9)$ & 0.7 \\ \hline\hline
     \end{tabular}
     \label{table:sim-params}
\end{table}

\begin{figure}[htb]
        \includegraphics[width=\columnwidth,clip=true]{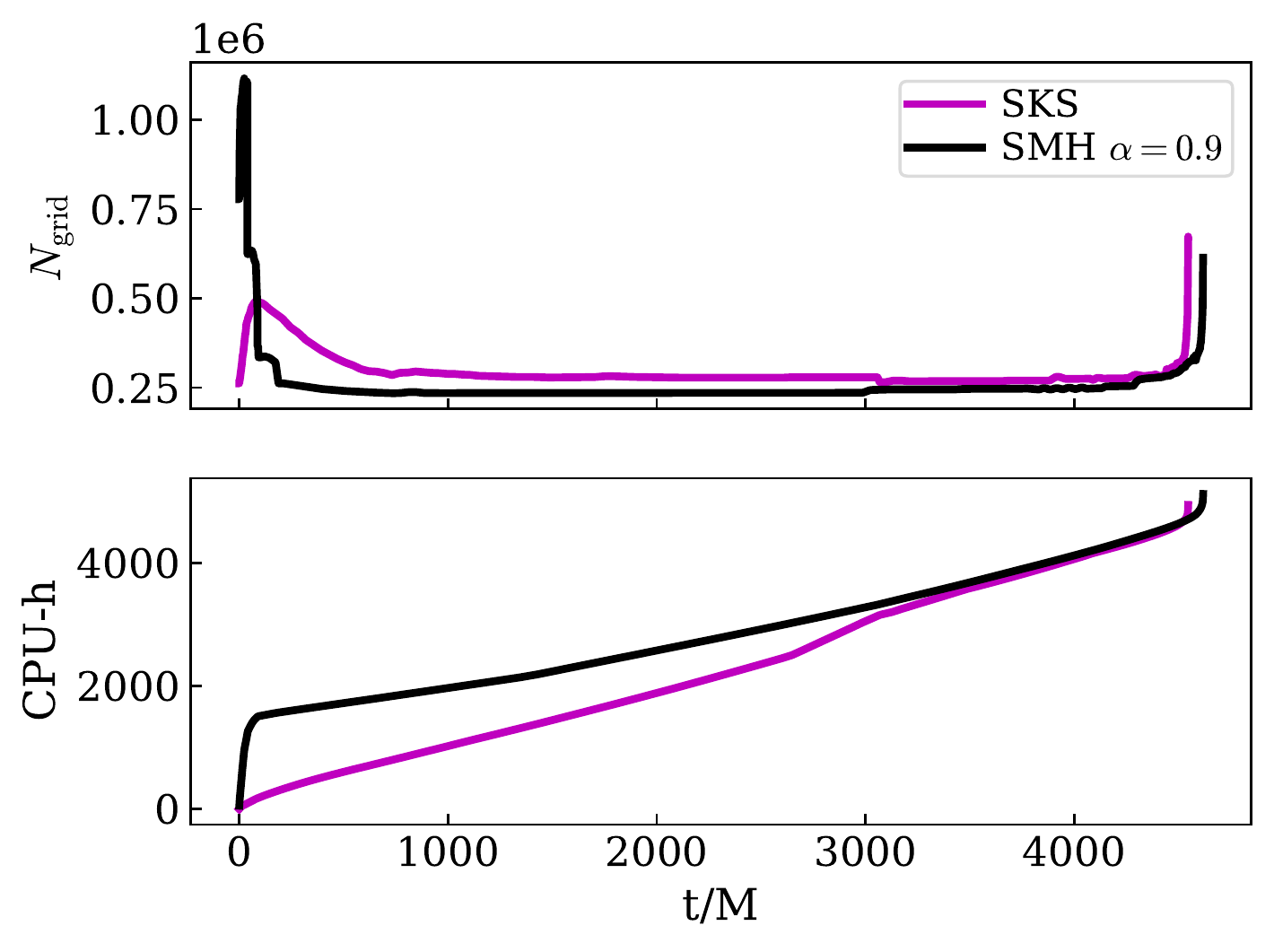} 
  \caption{
Computational efficiency of evolutions of SMH $(\alpha=0.9)$ and SKS
      initial data for Case I, with the highest resolution. The upper panel is the total number of grid points as a
      function of time. At the beginning, the SMH initial data requires many
      more grid points to meet the error tolerance.  As the gauge transition
      to damped harmonic gauge proceeds (on a time scale of $\sim 50\,M$),
      the BHs become less
      distorted, so
      AMR gradually drops points.  At the same time, several
      concentric spherical shells around each of the BHs are dropped, which
      leads to discontinuous jumps in the number of grid points. In the end,
      evolutions of SMH initial data has fewer collocation points than for SKS.
      The lower panel is the accumulated CPU hours versus time.  The SMH
      initial data is extremely slow at the beginning.  As the collocation
      points and subdomains are adjusted, it speeds up.  The total CPU hours
      for evolutions of both initial data
  sets are similar.}
 \label{fig:time}
\end{figure}

\begin{figure}[htb]
        \includegraphics[width=\columnwidth,height=6.2cm,clip=true]{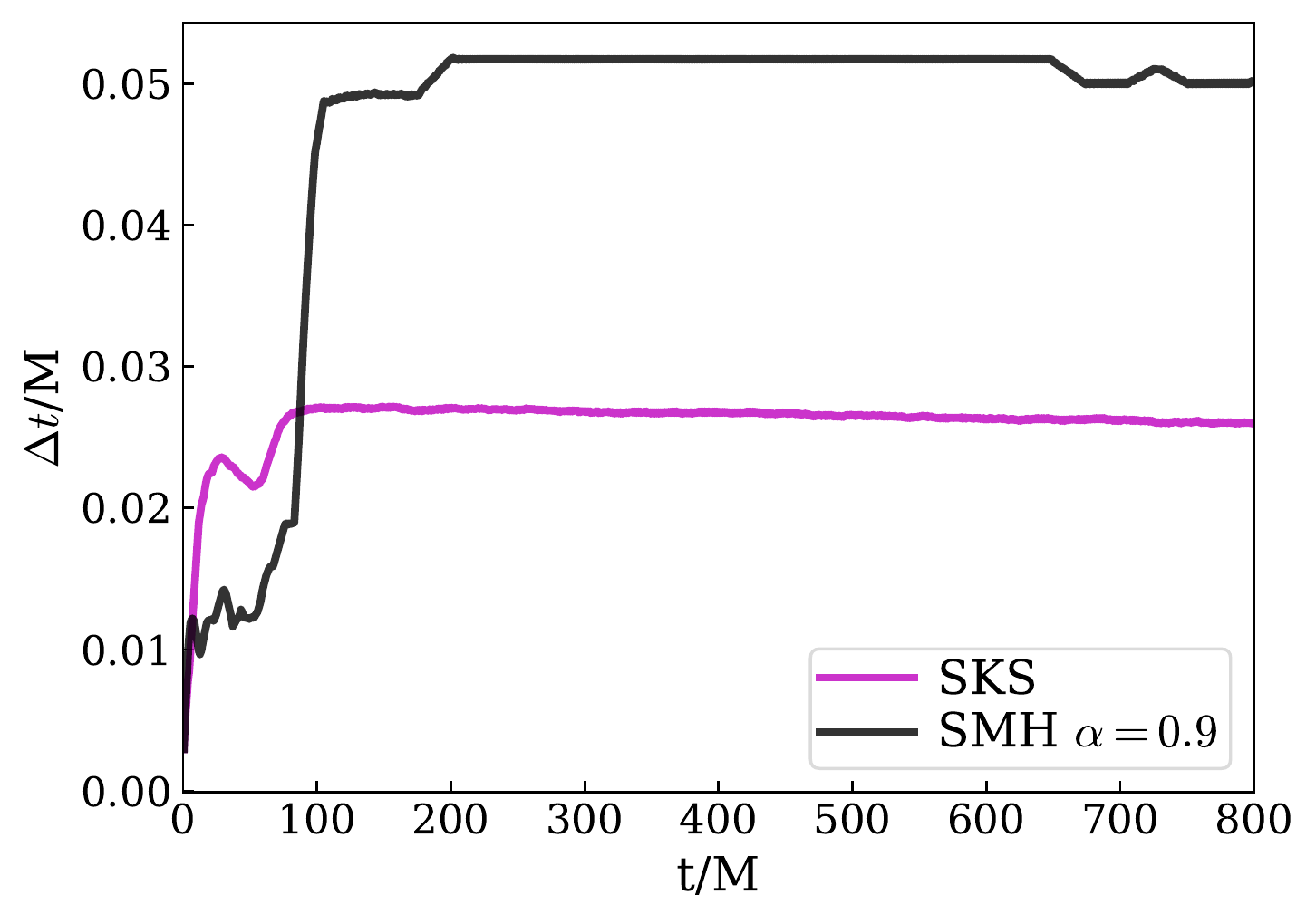} 
  \caption{The time step as a function of evolution time.
  The resolution is Lev 3. Initially, the time step for evolutions of
  SKS initial data
  is larger than for SMH. However, after several jumps
  due to the shell-dropping algorithm, SMH eventually
  has a larger time step than SKS.
}
 \label{fig:time-step}
\end{figure}

\subsection{Constraint violations and computational efficiency}

Figure \ref{fig:constraint} shows the evolution of the
volume-weighted generalized harmonic constraint energy $N_{\mathrm{volume}}$, which is given by [Eq.~(53) of Ref.~\cite{Lindblom:2005qh}]
\begin{align}
N_{\mathrm{volume}} = \sqrt{\frac{\int_V F(x)^2 d^3x}{\int_V d^3x}},
\end{align}
with $F(x)$ the generalized harmonic constraint energy at $x$.
 For the first $\sim25\,M$ of evolution,
the constraints of SMH are
much larger than those of SKS.
This is because BHs with SMH initial data are more distorted than SKS,
and the metric is
more difficult to resolve; however, the metric is much easier to resolve for SMH than SH (which is not shown because even constructing the initial data for SH
is problematic with a spin of $\chi=0.8$).
Furthermore, at slightly later times, constraints
 decrease rapidly.
During the junk stage ($t \lesssim 700\,M$), the constraints
for the evolution of
SMH initial data are smaller than those of SKS by an order of magnitude.
After the junk leaves the system, the evolution of
SMH initial data is still a
little bit better than that of SKS initial data, although constraints of
SKS and SMH become similar
at late times
($t\gtrsim3000\,M$ for Lev 3 and $t\gtrsim2000\,M$ for Lev 2).

During the junk stage
we make no attempt to resolve the junk oscillations, i.e., the AMR
algorithm is intentionally set to change the grid very
infrequently (and not at all in the wave zone)
during the junk stage of the evolution.  We do this because resolving junk is
computationally expensive and because the junk is not part of the physical
solution we care about. Accordingly, the SKS curves in
Figure~\ref{fig:constraint} are not well-resolved during the junk stage
and do not show good convergence.  However, we notice that the simulations
of SMH initial data are better resolved than for SKS, and they
converge with resolution even during the junk stage; convergence
during the junk stage
was also observed for SH with low-spin BHs \cite{Varma:2018sqd}.

The convergence plot looks slightly different when the
norm of the constraint energy is determined using a
pointwise $L^2$ norm over grid points rather than an integral
over the volume, as given by
\begin{align}
N_{\mathrm{pointwise}} = \sqrt{\frac{\sum\limits_{i=1}^{N} (F_i)^2}{N}},
\end{align}
where the subscript $i$ stands for the index of a
grid point, and $N$ is the total number of grid points.
The pointwise norm is shown in Fig.\ref{fig:constraint_L2}.
For the pointwise norm
the improvement of convergence of SMH over SKS
is not as good as for the
volume-weighted norm.
This is because the pointwise norm gives larger weight to the interior
regions near the BHs where there are more points, whereas the volume norm
gives larger weight to the exterior wave zone which covers more volume.
The difference beween Figs.~\ref{fig:constraint} and~\ref{fig:constraint_L2} illustrates that the improvement
of the constraints in the case of SMH mainly comes from the outer
region, where  the high-frequency components in the
waveforms are smaller
(i.e., less junk radiation). Figure \ref{fig:constraint_L2}
also shows that the pointwise norms ($L^2$ norm) for evolutions of
both initial data sets
become comparable much earlier than the volume norms
($t\sim 200\,M$). This
is because the pointwise norms are monitored by
AMR, and therefore
their values remain consistent with the numerical error tolerance
in AMR during the evolution as AMR makes changes to the grid resolution.

\begin{figure*}[htb]
        \includegraphics[width=\columnwidth,clip=true]{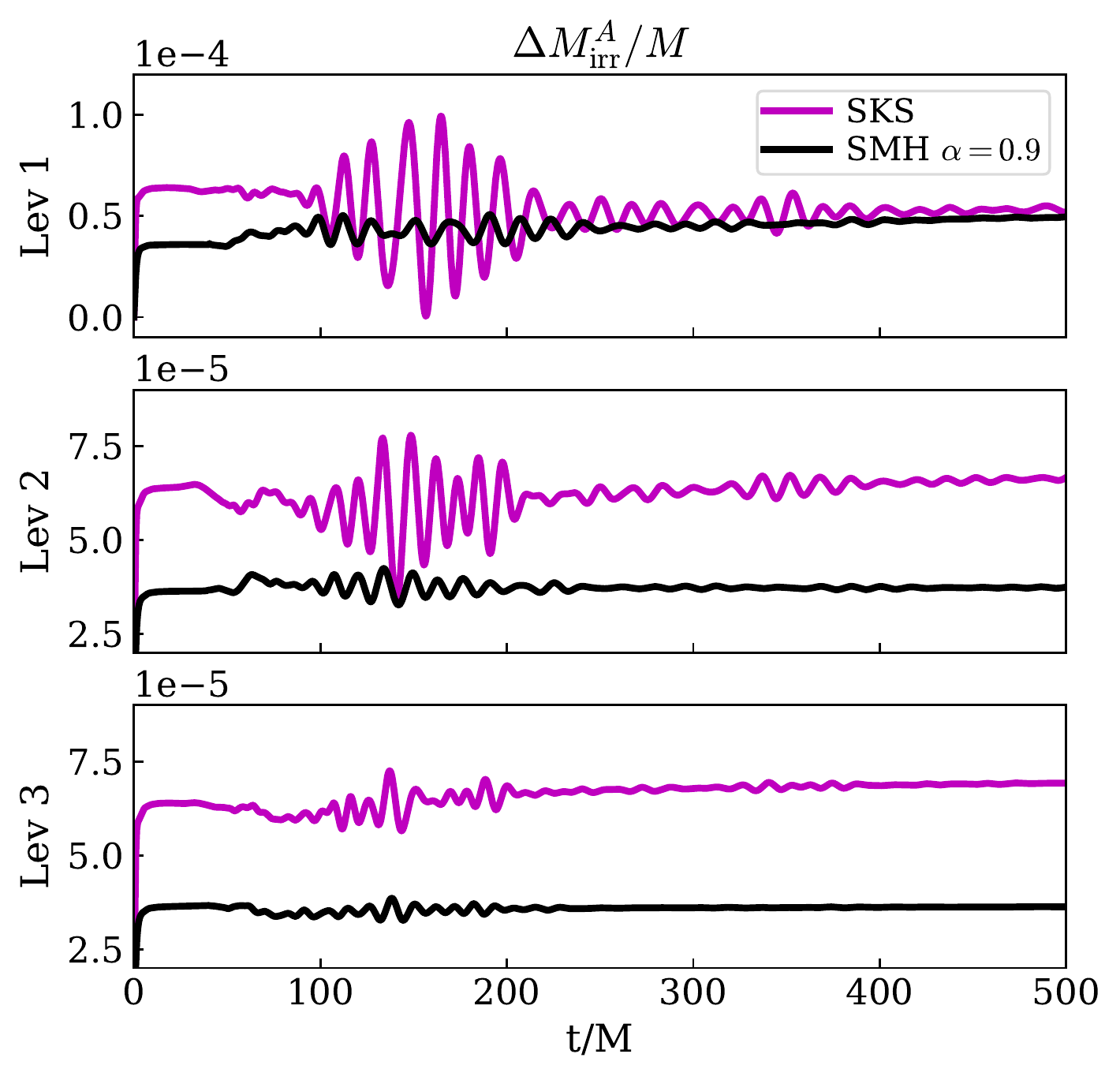} 
        \includegraphics[width=\columnwidth,clip=true]{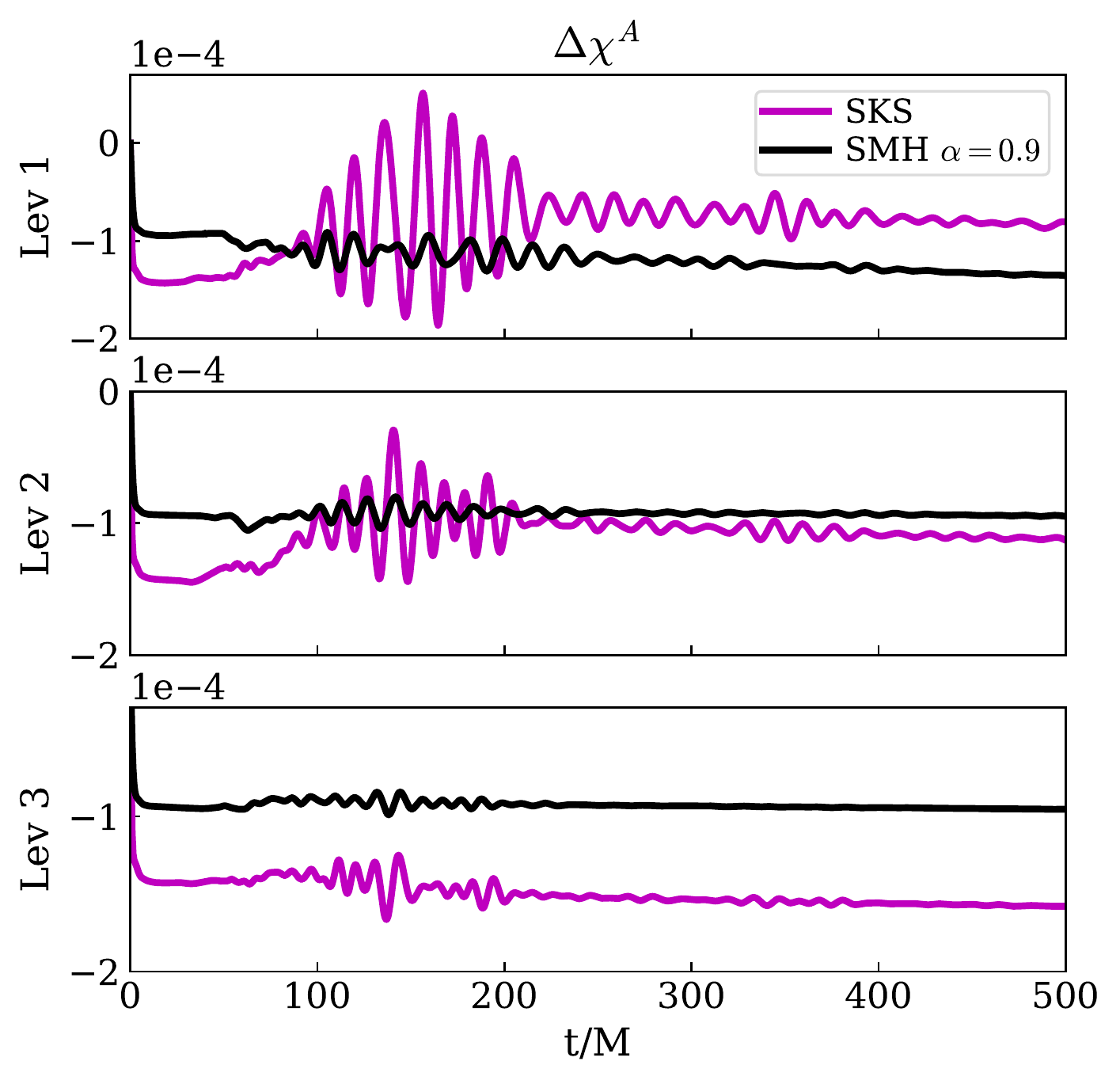} 
        \caption{
        The evolution of  irreducible mass (left) and dimensionless spin (right) of the first BH for Case I, with three resolutions.
        The quantities shown are deviations from their values at $t=0$.
        Evolutions of SMH initial data
 have fewer oscillations than SKS.
  Deviations of three parameters for both initial data sets are on the same order. 
}
 \label{fig:par-change}
\end{figure*}

To understand how
the computational
efficiency of the evolution depends on the initial data,
in Fig.\ \ref{fig:time} we show the total number of
grid points in the
computational domain as a function of time.
At the beginning, SMH needs many more points than SKS.
As the gauge gradually transforms to the damped harmonic gauge,
the BHs become less distorted and AMR decides to drop grid points.
During the evolution, there are two factors that mainly control
the number of grid points. One is AMR, which adjusts
grid points based on the numerical error tolerance.
The other one is the domain decomposition \cite{Ossokine:2015yla}.
SpEC splits the entire computation region into various subdomains.
In particular, there are a series of concentric spherical shells around each
BH.
The subdomain boundaries are fixed in the ``grid frame'', the frame
in which the BHs do not move, but these boundaries do move in
the ``inertial frame'', the frame in which the
BHs orbit and approach each other \cite{Scheel:2006gg}.
As the separation between the BHs decreases,
the inertial-frame widths of the subdomains between them decreases as well.
During the evolution, the inertial-frame
widths of the spherical shells are monitored.
Once one of the shells becomes sufficiently squeezed,
the algorithm drops one of the shells and redistributes the computational
domain.
In Ref.\ \cite{Varma:2018sqd}, the authors pointed out that evolutions of
SH initial data
are faster than for SKS initial data.
However, that statement is not
true at very early times, when SH
starts with more spherical shells and more grid points,
which leads to low speed. The evolution of SH initial data
then gradually speeds up after
several spherical shells are dropped, and eventually becomes
faster than the corresponding evolution of SKS data.
Our simulation here is similar.
In Fig.\ \ref{fig:time}, AMR modifies $N_\text{grid}$ smoothly,
while the discontinuous jump is caused by the shell-dropping algorithm.
For each BH, we have six spherical shells initially. However, four of
them are dropped during the first $\sim200\,M$.
In the end, the number
of grid points for evolutions of SMH is smaller than
for evolutions of SKS. This not only improves
the computational efficiency
of each time step, but also increases the
time step $\Delta t$ allowed by the Courant
limit
($\Delta t\sim N^{-2}_\text{grid}$).
As shown in Fig.\ \ref{fig:time-step}, the time step for SMH
jumps several times
because of the shell-dropping algorithm.
In the end, $\Delta t$ for SMH is larger than the one for SKS.
Both $N_\text{grid}$ and $\Delta t$ contributes to the high speed
of evolutions of
SH and SMH initial data. And we have checked that
the increase of $\Delta t$
plays the major role in the speed increase. 

The bottom
panel of Fig.~\ref{fig:time} shows the
accumulated CPU hours of the simulation. At first, the evolution of
SMH is extremely slow. Once several shells are dropped, the simulation
gradually speeds up. This suggests
that both SH and SMH initial data
start with more shells than necessary.
Therefore, it  might be possible to
further
improve the computational efficiency solely by 
reducing the number of shells.

\begin{figure*}[htb]
       \includegraphics[width=\textwidth,clip=true]{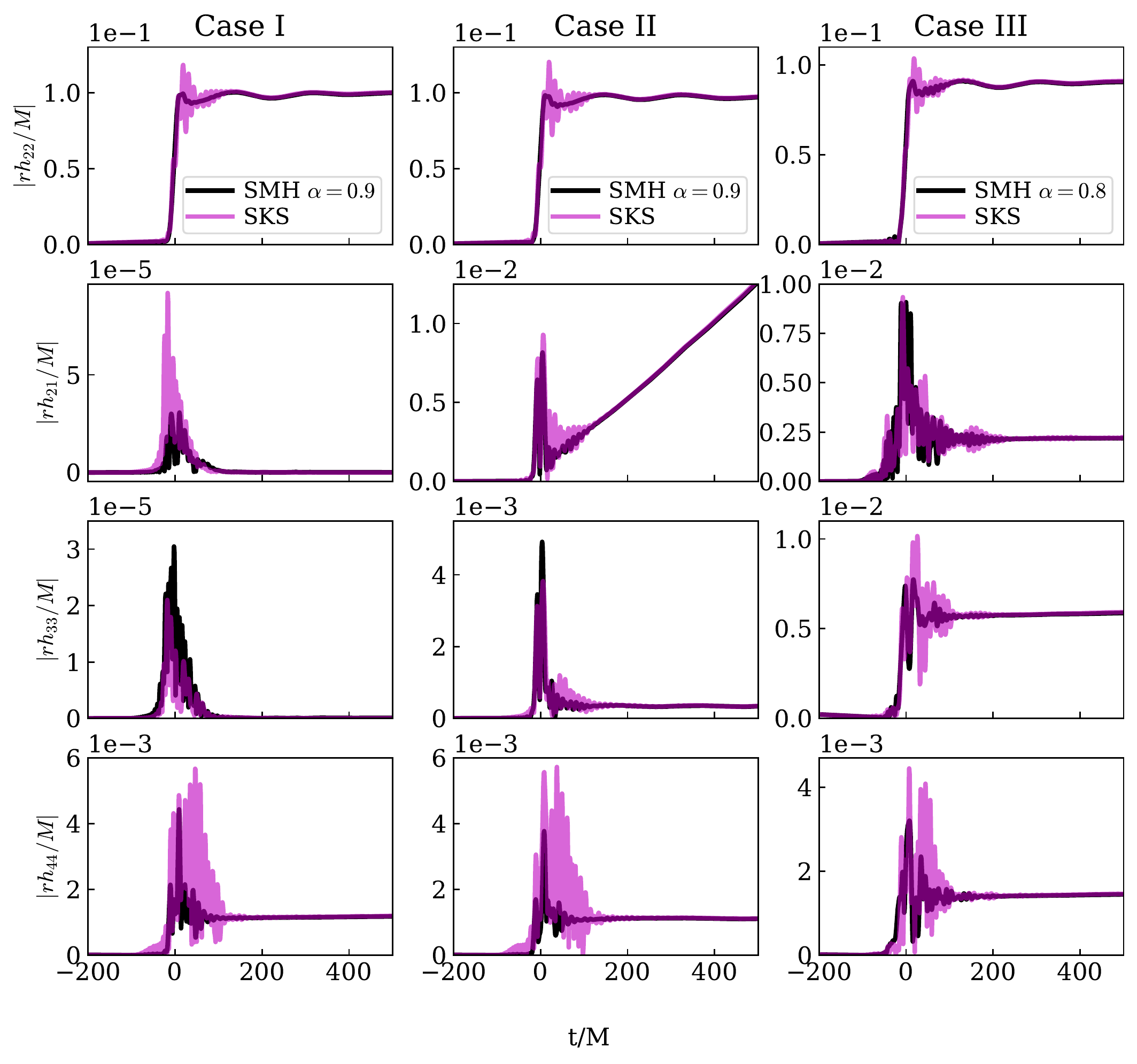} 
  \caption{Mode amplitudes of waveforms for Case I, II and III with the highest resolution.
  Columns correspond to three cases, and rows are for different modes. For
  SMH initial data, we pick $\alpha=0.9$ for Case I and II, and
  $\alpha=0.8$ for Case III.
  Note that the linear growth of $h_{21}$ for Case II
   appears because
  only the initial part of the waveform is shown. Over the entire evolution, the mode is oscillatory. In general, the junk radiation of SMH initial data leaves the system faster. It is also
  smaller than the junk radiation of SKS for most of the modes.
  However, there are some modes, such as $h_{33}$, that have the same peak as
  SKS.  }
 \label{fig:waveform}
\end{figure*}

\subsection{Junk radiation and changes in parameters}
Since the BHs in the initial data are not in true
quasi-equilibrium, the masses and spins of BHs relax once the evolution begins, resulting in slight deviations from their initial values.
In Figure \ref{fig:par-change}, we show the change of irreducible mass $\Delta M_\text{irr}(t)=|M_\text{irr}(t)-M_\text{irr}(t=0)|$
and the change of spin $\Delta\chi(t)=|\chi(t)-\chi(t=0)|$ as functions of
time, for three resolutions.
We can see the variations are on the same order for both SMH and SKS
initial data, but SMH has smaller oscillations.
With the highest resolution,
the deviation of SMH is smaller by a factor of $\sim1.5-2$. 

\begin{figure}[htb]
        \includegraphics[width=\columnwidth,height=9.5cm,clip=true]{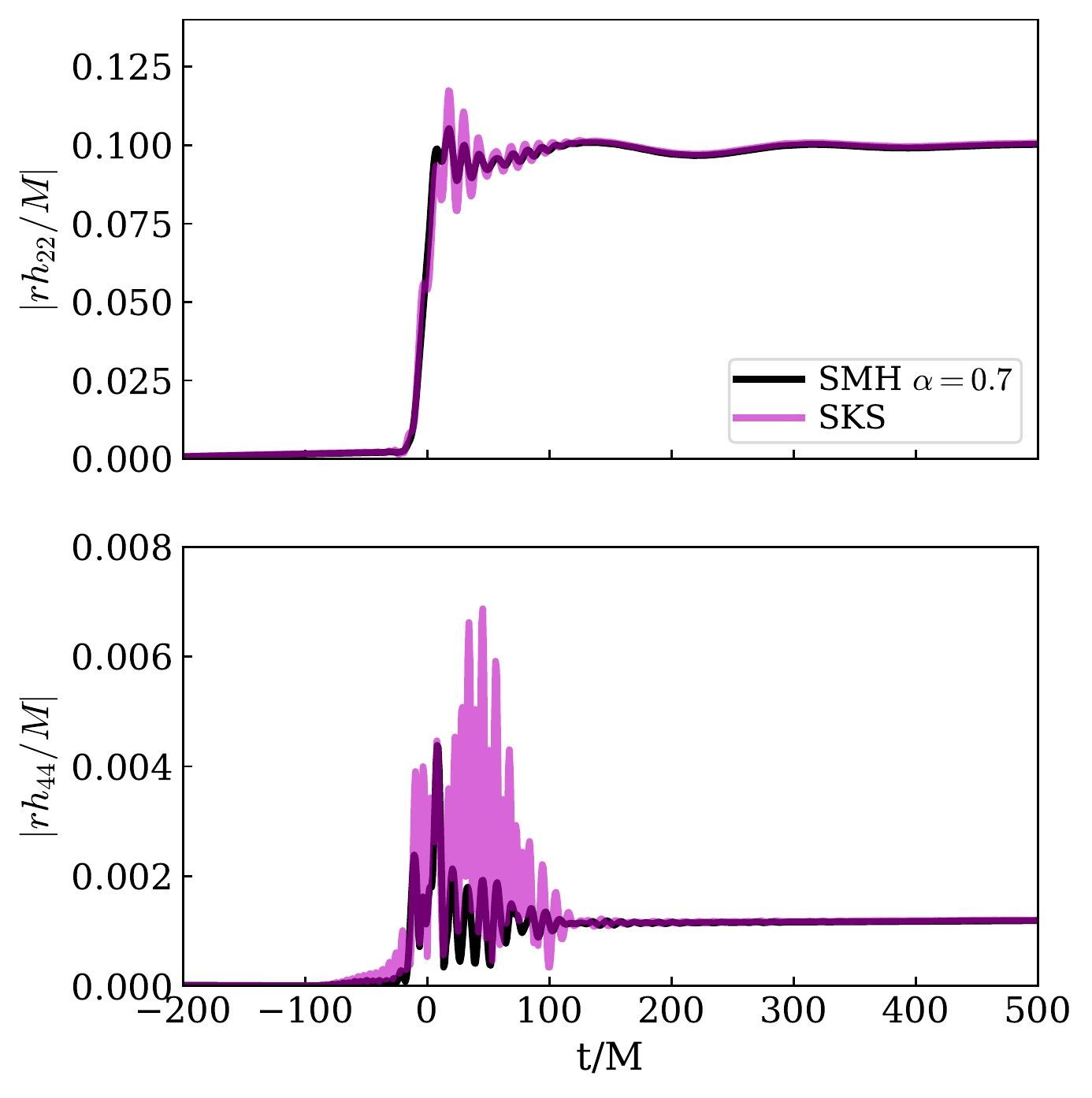} 
  \caption{ The $h_{22}$ and $h_{44}$ modes for the highest resolution
   of Case IV, an equal-mass BBH system with larger spins.
   The spins for both BHs are $(0,0,0.9)$,
   which we have not been able to run with SH initial data.
  We can still see that the junk radiation for SMH is less than SKS.
}
 \label{fig:waveform-high-spin}
\end{figure}

To study the junk radiation in the waveform, in Fig.\ \ref{fig:waveform} we plot the amplitudes of different spin weighted spherical harmonic modes $h_{lm}$, for Case I, II and III listed in Table \ref{table:sim-params} (Case IV will be discussed later). Note that the linear growth of $h_{21}$ for Case II appears because only the initial part of the waveform is shown;
over the entire evolution, the mode is oscillatory. 

We can see that the junk radiation of evolutions of SMH initial data is less
than for SKS for most of the modes. In general, the junk radiation leaves the
system faster for SMH initial data than for SKS. However, the decrease of junk
radiation for SMH
is not as significant as SH for low-spin BHs~\cite{Varma:2018sqd}.
Some modes of SMH initial data, such as $h_{33}$, are similar to SKS.
Comparing Cases II and III, we note
that the junk radiation of $\alpha=0.8$ SMH is larger than that of $\alpha=0.9$
, presumably
because $\alpha=0.8$ deviates more from SH initial data ($\alpha=1$).
Note that Case II has similar junk radiation as Case III
when both cases are evolved from SKS initial data; this
suggests that the difference in junk radiation between Cases II and III
seen in Figure~\ref{fig:waveform} is probably not due to
differences in parameters like the mass ratio.

\begin{figure}[htb]
        \includegraphics[width=\columnwidth,height=6.8cm,clip=true]{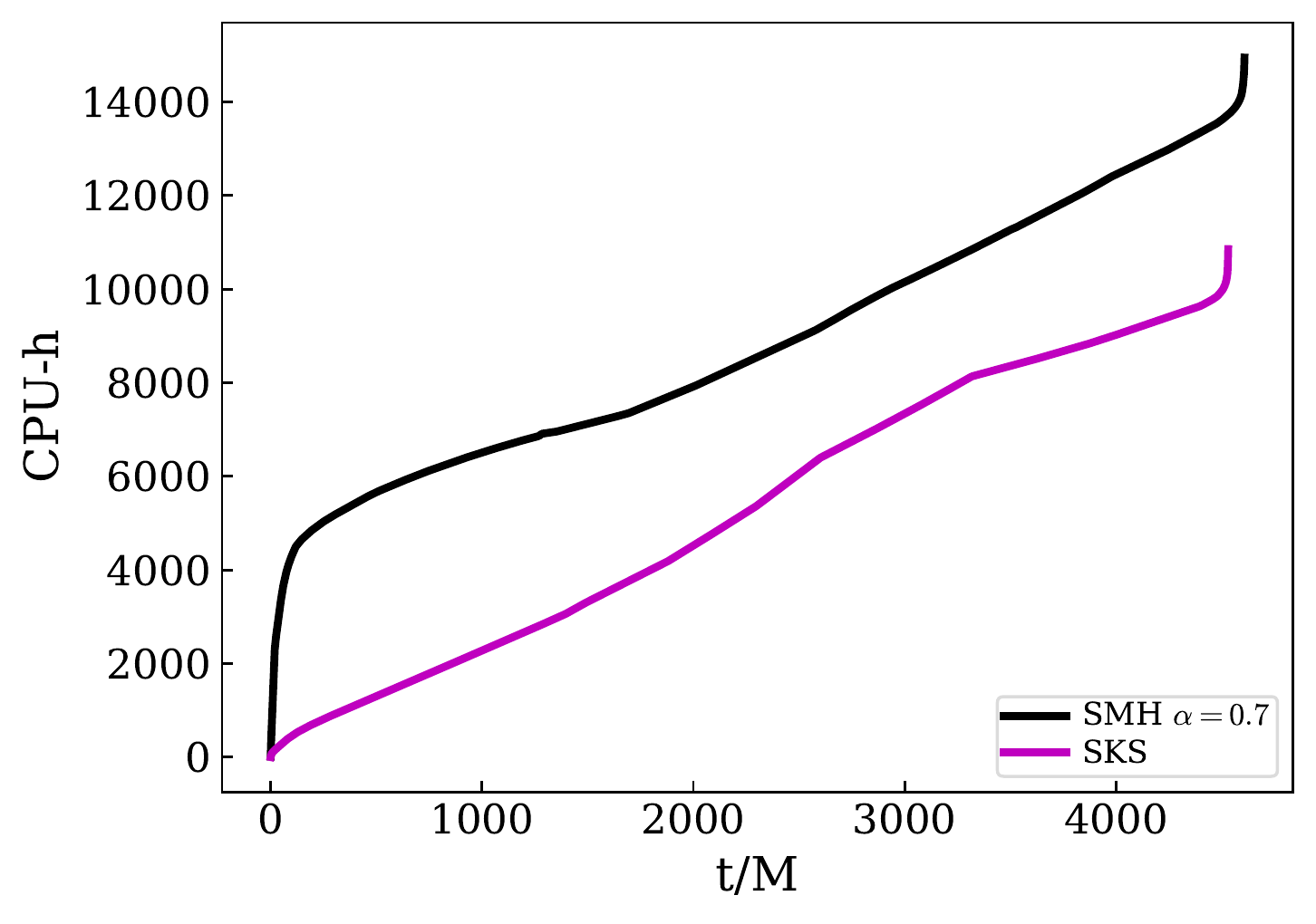} 
  \caption{
  The accumulated CPU hours for evolutions of SMH and SKS initial
      data as functions of time. The BBH system is
       Case IV, and we plot results for the highest resolution.
      The initial computational efficiency of SMH initial data is much lower
      than for SKS, but after a short time both evolutions proceed
      at the same number of CPU hours per simulation time. 
}
 \label{fig:time-high-spin}
\end{figure}

For Case IV, 
a BBH system with dimensionless spins 0.9, we need to
decrease $\alpha$ to 0.7, since for that large of spin $\alpha=0.8$ requires too high resolution and sometimes the initial data solver doesn't converge.
To speed up the evolution, we start the SMH initial data with
fewer spherical shells around each BH than the standard choice made by SpEC.
The comparison of the waveform
is in Fig.~\ref{fig:waveform-high-spin}, where we show only $h_{22}$ and
$h_{44}$. We can see the junk radiation for SMH initial data is still less than
for SKS. But the improvement is not as good as other cases. For modes other than $h_{22}$ and
$h_{44}$ , we
do not see improvements. The main reason
appears to be that $\alpha=0.7$ deviates too much
from $\alpha=1$, so that the benefit of SH initial data is reduced. In
addition, in Fig.~\ref{fig:time-high-spin} we compare the accumulated CPU hours for evolutions of both
initial data sets.
We can see the initial computational
efficiency for SMH initial data is much lower, but it gradually catches
up after several shells are dropped. For evolutions of only a few orbits, the
expense of evolving SMH initial data may not be worth the extra computational
cost.  But for evolutions of many orbits, the extra cost at the beginning of
the evolution will be comparatively small.

In most of the evolutions shown here, shortly after
the beginning of the simulation several
spherical shells around each BH are dropped, leading to a smaller
number of grid points, a larger time step, and overall greater computational
efficiency.  However, for a general evolution, we are not always
`lucky' enough to gain this efficiency, since the current algorithm for
dropping spherical shells aims only to avoid narrow shells
rather than to speed up the
simulation.
To improve the computational efficiency for all simulations,
we could start with fewer spherical shells at $t=0$. However, the benefit
of this change is
limited without changing the shell-dropping algorithm.
One workaround is to use smaller
$\alpha$, which speeds up the simulation,
but if  $\alpha$
deviates too much from $\alpha=1$, we cannot have less junk
radiation.
Therefore, we suggest that the algorithm that divides
the domain in to subdomains
should be modified to account for
computational efficiency during the evolution, or a better algorithm
should be developed to
initialize subdomains. Given such future algorithmic
improvements, we could
potentially run high-spin BBH evolution with larger $\alpha$, which can lead to
less junk radiation.

\section{Conclusion} 
\label{sec:conclusion}

In this paper, we extended SH initial data~\cite{Varma:2018sqd} to
higher-spin BBHs by introducing a class of spatial coordinate
 systems that represent a time-independent
slicing of a single Kerr black hole and are
characterized by a continuous parameter $\alpha$.
This coordinate representation of Kerr is used to supply
free data for the initial-value problem for BBH systems; we call the
resulting initial-value solution SMH initial data. The harmonic $(\alpha=1)$ and KS
$(\alpha=0)$ coordinate
representations of Kerr are only two special cases of our
new representation.  The coordinate shape of the horizon
becomes less spherical and more distorted for larger $\alpha$.
Therefore for high-spin BHs, we pick $\alpha<1$ to decrease the
distortion and ease requirements on very high resolution during the
BBH simulation. At the same time, $\alpha$ should be close to 1 so that
SMH initial data still has the desirable properties of SH initial data
as shown in Ref.~\cite{Varma:2018sqd}, such as less junk radiation. We
have tested that for SMH initial data with $\alpha=0$, i.e, harmonic
time slicing with KS spatial coordinates, there is more junk radiation
than for SKS initial data.

We have evolved four
BBH systems with dimensionless spins 0.8 or 0.9
starting from SMH initial data with $\alpha$ between 0.7 and
0.9, and we compared with evolutions of the
same system starting from SKS initial data.  The first three cases, all with dimensionless spins 0.8, represent
different situations: a non-precessing system with equal masses, a
precessing system with random spin directions, and a non-precessing
system with unequal masses. In general, the junk radiation of SMH
initial data leaves the system faster than that of
SKS. For most gravitational
wave modes, the SMH initial data leads to less junk radiation. The
exceptions, like the $h_{33}$ mode and the $h_{21}$ mode for Case III,
have bursts with amplitudes similar to SKS. Furthermore, $\alpha=0.8$
SMH has more junk radiation than $\alpha=0.9$.

Using Case I as an example, we also studied other
properties of the evolution,
including constraint violations, computational efficiency,
and changes in parameters.
We found the values of the volume-weighted constraints for SMH initial
data are smaller than those of SKS
by factors of 10.
Furthermore, the volume-weighted constraints of SMH initial data converge
with resolution during the junk stage. However, $L^2$-norm constraints do not
have such convergence.  Therefore, the benefit is mainly from the outer
regions, where there is less junk radiation.

At the beginning of the
evolution for Case I,
SMH requires more collocation points than SKS to reach the error
tolerance because the horizon is distorted, hence it
proceeds more slowly.  At later times, SKS and SMH run at approximately
the same rate, after both the computational efficiency on each time slice and
the size of the time step increase for the SMH case.

For Case IV, which has
BHs with dimensionless spin $0.9$, we found that we needed to decrease
$\alpha$ to $0.7$. We simulated an equal-mass BBH system with equal
dimensionless spins $\bm{\chi}_{1,2}=(0,0,0.9)$ and compared $h_{22}$ and $h_{44}$ for
both SMH and SKS initial data sets.  Junk radiation for SMH is still less than
for SKS, but the improvement is not as good as the case of lower spin. The
comparison of CPU hours for these two cases show that the initial computation
efficiency for SMH initial data is much lower. But it gradually becomes the
same as SKS after several shells are dropped. 

We also found that the algorithm for choosing the number and sizes of
subdomains in SpEC could use some improvement, particularly for the
initial choice of subdomains and the early stages of the evolution.
In most simulations but not all, AMR eventually chooses a subdomain
distribution that increases computational efficiency.  Some improvements
can be gained by simply starting with fewer spherical shells around each
BH, but we find that the effects of this change are limited.
Therefore, the evolution of SMH initial data for high-spin BBH will
benefit from either an algorithm to adjust subdomain sizes based on
computational efficiency during the evolution, or a better algorithm to
initialize subdomains. Those algorithmic improvements
could allow us to run high-spin BBH evolutions with larger $\alpha$,
which can give rise to less junk radiation.


\begin{acknowledgments}
We want to thank Maria Okounkova, Saul Teukolsky and Harald Pfeiffer for useful
discussions. 
M.G.\ is supported in part by NSF Grant PHY-1912081 at Cornell.
M.S.\ and S.M.\ are supported by NSF Grants No.
PHY-2011961, PHY-2011968, and OAC-1931266 at Caltech. 
V.V.\ is supported by a Klarman Fellowship at Cornell.
M.G. and V.V. were supported by NSF Grants No. PHY-170212 and PHY-1708213 at
Caltech.
S.M, M.G, M.S, and V.V. are supported by the Sherman Fairchild Foundation.
The computations presented here were conducted on the
Caltech High Performance Cluster, partially supported by a grant from
the Gordon and Betty Moore Foundation. 
This work was supported in part
by NSF Grants PHY-1912081 and
OAC-1931280 at Cornell.
\end{acknowledgments}


\appendix
\section{Details of MH coordinates}
\label{app:metric}

For a Kerr BH with an arbitrary spin vector $\bm{a}$, the transformations between KS spatial coordinates and MH spatial coordinates are given by
\begin{align}
\bm{x}_\ks=&\frac{a^2+r(r-\alpha M)}{a^2+(r-\alpha M)^2}\bm{x}_\mh\nonumber\\
+&\frac{\alpha M}{a^2+(r-\alpha M)^2}(\bm{x}_\mh\times\bm{a})\nonumber\\
+&(\bm{x}_\mh\cdot\bm{a})\bm{a}\frac{\alpha M}{(r-\alpha M)[a^2+(r-\alpha M)^2]}, 
\end{align}
where $a^2=\bm{a}\cdot\bm{a}$, and $r$ is the radial Boyer-Lindquist coordinate. For
$\alpha=0$, we have $\bm{x}_\ks=\bm{x}_\mh$, i.e., the identity transformation.
The Jacobian $C^{ij}_\mh=\partial x_\ks^i/\partial x_\mh^j$ between KS and MH coordinates is given by\footnote{Here we do not distinguish upper and lower indices of a tensor in a Euclidean space.}
\begin{align}
C^{ij}_\mh &= \frac{a^2+r(r-\alpha M)}{a^2+(r-\alpha M)^2}\delta^{ij}
+\frac{\alpha M}{a^2+(r-\alpha M)^2}a_k\epsilon^{ijk}\nonumber\\
&+a^ia^j\frac{\alpha M}{(r-\alpha M)[a^2+(r-\alpha M)^2]} \notag \\
&+\frac{M\alpha[a^2-(r-M\alpha)^2]}{[a^2+(r-M\alpha)^2]^2}x_\mh^i\partial^jr
\nonumber\\
&-\frac{2M\alpha(r-M\alpha)}{(a^2+(r-M\alpha)^2)^2}x^\mh_ma_k\epsilon^{imk}\partial^jr\nonumber\\
&-x_\mh^ma_ma^i\partial^jr\frac{M\alpha[a^2+3(r-M\alpha)^2]}{[a^2+(r-M\alpha)^2]^2(r-\alpha M)^2},
\end{align}
where $\epsilon^{ijk}$ is the Levi-Civita symbol, $\delta^{ij}$ is the Kronecker delta, and the Einstein summation convention is used.
For $\alpha=1$, $C^{ij}_\mh$ becomes $C^{ij}$ defined in
Sec.~\ref{sec:hk}. By differentiating Eq.\ (\ref{r-x-mh}), we have
\begin{align}
\partial_i r=\frac{x_i^\mh+(\bm{a}\cdot\bm{x}^\mh) a_i/(r-\alpha M)^2}{2(r-\alpha M)\left[1-\frac{\bm{x}^\mh\cdot\bm{x}^\mh-a^2}{2(r-\alpha M)^2}\right]}.
\end{align}
With MH coordinates, the null covariant vector $l$ in Eq.\ (\ref{metric-ks}) can be written as
\begin{align}
l&=\left(dt_\mh+\frac{2M}{r-r_-}dr\right)\nonumber\\
&+\frac{\rma \bm{x}_\mh-\bm{a}\times\bm{x}_\mh+(\bm{a}\cdot\bm{x}_\mh)\bm{a}/\rma}{\rma^2+a^2}\cdot d\bm{x}_\mh,
\end{align}
where the first bracket corresponds to $dt_\ks$ [see Eq.\ (\ref{time-hk-ks}), with $t_\mh=t_\hk$]. The
scalar function $H$ in Eq.\ (\ref{metric-ks}) is given by
\begin{align}
H=\frac{Mr\rma^2}{r^2\rma^2+(\bm{a}\cdot\bm{x}_\mh)^2}.
\end{align}
In addition, the lapse function $N$ and the shift vector $N^i$ in MH coordinates are given by
\begin{align}
&N^{-2}=1+\frac{2M(r-\alpha M)^2}{r^2(r-\alpha M)^2+\adx^2}\frac{r^2+(r+2M)r_+}{r-r_-}, \\
&N^i=N^r l^i+N^\phi\frac{a_jx^\mh_k\epsilon^{jki}}{a},
\end{align}
with
\begin{align}
&N^r=N^2\frac{2Mr_+}{\rho^2},\quad N^\phi=-N^2\frac{a}{\rho^2}\frac{2M}{r-r_-}, \\
&\rho^2=r^2+a^2\cos^2\theta=r^2+\frac{\adx^2}{(r-\alpha M)^2}, \quad l^i=l_i,
\end{align}
where $\theta$ is the polar Boyer-Lindquist coordinate, and $l_i$ is the spatial component of the null covariant vector $l$.

%

\bibliography{References}

\end{document}